\documentclass[prb,twocolumn,showpacs,preprintnumbers,amsmath,amssymb]{revtex4}
\usepackage{graphicx}
\usepackage{dcolumn}
\usepackage{bm}

\begin{document}

\title{Properties of KCo$_2$As$_2$ and Alloys with Fe and Ru: Density 
Functional Calculations}

\author{D.J. Singh}

\affiliation{Materials Science and Technology Division,
Oak Ridge National Laboratory, Oak Ridge, Tennessee 37831-6114}

\date{\today}

\begin{abstract}
Electronic structure calculations are presented for KCo$_2$As$_2$
and alloys with KFe$_2$As$_2$ and KRu$_2$As$_2$. These materials show
electronic structures characteristic of coherent alloys, with
a similar Fermi surface structure to that of the Fe-based superconductors,
when the $d$ electron count is near six per transition metal.
However, they are less magnetic than the corresponding Fe compounds.
These results are discussed in relation to superconductivity.
\end{abstract}

\pacs{71.20.Lp,74.70.Dd}

\maketitle

\section{introduction}

A remarkable feature of the high temperature superconductivity in iron
compounds is the chemical richness of this phenomenon.
Besides the oxypnictides, \cite{kamihara-a}
superconductivity has been discovered in related ThCr$_2$Si$_2$ structure
materials (prototype BaFe$_2$As$_2$),
\cite{rotter-sc}
LiFeAs and related compounds, \cite{wang-lifeas}
perovskite - pnictide intergrowth (Sr$_4$Sc$_2$Fe$_2$P$_2$O$_6$) compounds,
\cite{ogino}
and remarkably FeSe, \cite{hsu,mizuguchi} which contains no pnictogens.

In fact, the discovery of superconductivity in FeSe was an important
development, since it showed that pnictogen coordinated Fe was
not an essential part of Fe-based superconductivity.
Nonetheless, in spite of this diverse chemistry the Fe-based superconductors
do share a number of common features.
In particular, they are all based on divalent iron square planes, with 
tetrahedral coordination by pnictogens or chalcogens. This leads to
a characteristic electronic structure, in which Fe occurs as
Fe$^{2+}$ and the band structure near the Fermi energy is derived
from Fe $d$ states, with only modest hybridization of ligand $p$ states.
\cite{singh-du}

The band structures show relatively small disconnected
Fermi surfaces, in particular hole sections near the 2D zone center
and electron sections at the zone corner. On the other hand, despite
the low carrier density, the density of states, $N(E_F)$ of these
compounds is high ($\sim$ 2 eV$^{-1}$ per Fe), placing them near
itinerant magnetism. An important feature is that hole and electron
sheets of Fermi surface in these compounds are nested,
\cite{mazin,kuroki}
which leads
to a spin density wave (SDW) instability in most of the undoped materials,
and which in any case places the materials near an SDW.
\cite{cruz,dong-sdw,rotter}
The association between the SDW, the related nesting
and superconductivity is much discussed
and is one of the central issues in the physics of these materials.

A second much discussed issue is the relationship of the Fe-based
high-$T_c$ compounds with the cuprates. Besides the obvious distinctions,
such as multi-orbital vs. single orbital band structures, 
a key point is that the Fe-based materials appear to be much more
like conventional metallic compounds in their behavior. For example,
a prominent Fermi edge is seen in spectroscopies, with little
evidence for Hubbard bands or other characteristic features of a
Mott-Hubbard system.
\cite{lu,kurmaev}
Perhaps related to this, and in strong contrast with cuprates, where
the correlated atomic physics of Cu$^{2+}$ is thought to be crucial,
superconductivity can be induced in the Fe compounds by alloying with
other transition elements, in particular Co, Ni, Ru, and other $4d$ and
$5d$ elements.
\cite{sefat-co,leithe-jaspar,paulraj,han}

This raises the question of how essential Fe is for Fe-based superconductivity.
Specifically, is the phenomenon the result of a specific band structure
with itinerant magnetism, or is correlated atomic physics specifically
associated with Fe$^{2+}$ essential?
As a step towards addressing this question, we focus on the ThCr$_2$Si$_2$
structure, which accommodates an exceptionally large variety of compositions,
\cite{pearson,pearson2,just}
and on the possible replacement of Fe$^{2+}$ by Co$^{3+}$.
We note that in oxides where correlated behavior may play an important
role, Co$^{3+}$ compounds show very different behaviors from Fe$^{2+}$
compounds (e.g. LaCoO$_3$ or NaCoO$_2$ vs. FeO).
In particular,
we use electronic structure calculations for KCo$_2$As$_2$,
and pseudobinary alloys with KFe$_2$As$_2$ and KRu$_2$As$_2$
to show that the specific band structure features of the Fe superconductors
can be realized in phases with little or no Fe. From an electronic
point of view, we find that KFe$_y$Co$_{2-y}$As$_2$ for Fe concentrations
$y/2$ well below the nearest neighbor square lattice percolation
threshold of 0.5927, \cite{lee}
is very closely
analogous to superconducting BaFe$_{2-x}$Co$_x$As$_2$,
and furthermore that
KRu$_y$Co$_{2-y}$As$_2$ is similar but further from magnetism.
We note that KCo$_2$As$_2$ is a known compound,
\cite{rozsa},
as are KFe$_2$As$_2$ and KRu$_2$As$_2$.
\cite{rozsa,wenz}

\section{approach}

The calculations shown here were performed within the local density
approximation (LDA) with the general potential linearized
augmented planewave (LAPW) method, including local orbitals.
\cite{singh-book}
Well converged zone samplings and
basis sets, including local orbitals for semi-core states and to
relax linearization errors, \cite{singh-lo}
were used with LAPW sphere radii of 2.2 Bohr
for K, and 2.1 Bohr for the other elements.
In all cases, we used the experimental lattice parameters for
the end-point compounds, and linearly averaged values derived from them
for the mixed compositions.
The internal coordinates $z_{As}$ were relaxed
by energy minimization for ordered cells, while in the virtual crystal
calculations the values of $z_{As}$ were interpolated between the
compositions with ordered cells ($y$=0, 1 and 2).
The details are similar to those in our previous calculations for
ThCr$_2$Si$_2$ pnictides.
\cite{singh-bfa,sefat-baco}

\section{KCo$_2$As$_2$}

\begin{figure}[tbp]
\vspace{0.25cm}
\includegraphics[height=0.96\columnwidth,angle=270]{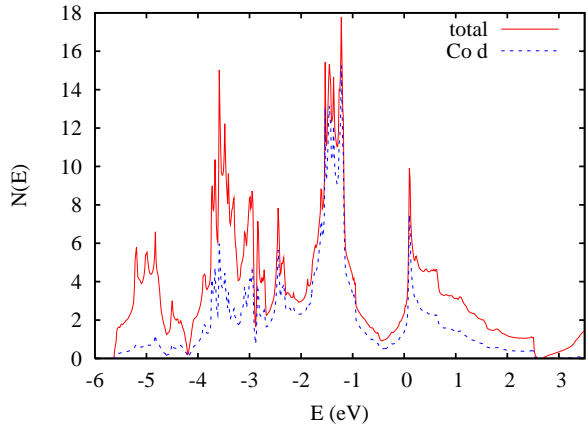}
\caption{(Color online)
Calculated density of states for KCo$_2$As$_2$ and Co $d$ contribution
as obtained by projection onto the Co LAPW spheres, radius 2.1 Bohr.}
\label{dos-C0R}
\end{figure}

\begin{figure}[tbp]
\vspace{0.25cm}
\includegraphics[height=0.96\columnwidth,angle=270]{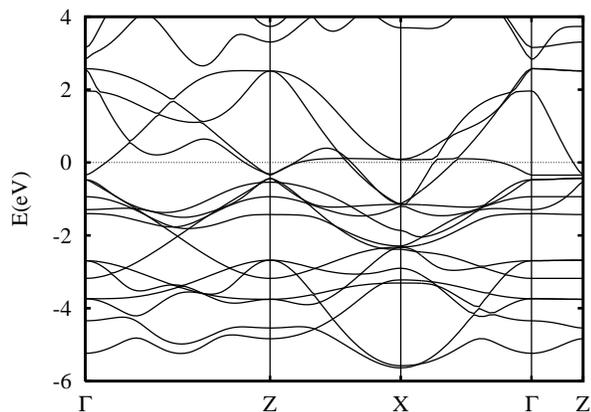}
\caption{
Calculated band structure of KCo$_2$As$_2$ showing lines in the
bct basal plane and perpendicular to it.
The body centered tetragonal reciprocal lattice vectors are
($2\pi/a,0,-2\pi/c$), ($0,2\pi/a,-2\pi/c$), and ($0,0,4\pi/c$).
In terms of these,
the long $\Gamma$-$Z$ direction is from 
(0,0,0) to (1,0,1/2) in the body centered tetragonal zone, while
the short $\Gamma$-$Z$ direction is from
(0,0,0) to (0,0,1/2). $X$ denotes the zone boundary (1/2,1/2,1/2) point.
A two dimensional band structure
would show no dispersion along the short $\Gamma$-$Z$
direction and would be symmetric about the mid-point of the long
$\Gamma$-$Z$ direction.}
\label{bands-C0R}
\end{figure}

The calculated electronic density of states (DOS)
and Co $d$ projection for KCo$_2$As$_2$ is shown
in Fig. \ref{dos-C0R} and the band structure in Fig. \ref{bands-C0R}.
These are as obtained in a non-spin polarized calculation
with the LDA relaxed As position $z_{\rm As}$=0.3452 and the experimental
lattice parameters. \cite{rozsa}
The basic shape of the DOS is very similar to that of the Fe superconductors,
except that the position of the Fermi energy is higher reflecting the
higher electron count.
In particular the DOS shows the characteristic $d$ derived density of states
with a pseudogap at a $d$ electron count of six per Co and
modest Co $d$ -- As $p$ hybridization.
However, the states where the Fermi energy, $E_F$
is positioned, i.e. at the lower edge of the upper DOS peak, are more
hybridized with As than the states on the edge of the lower peak where
$E_F$ is positioned in the Fe based materials.
The value at the $E_F$ is $N(E_F)$=2.4 eV$^{-1}$ per
formula unit both spins, of which 59\%
derives from the Co $d$ projection.
The reduced $d$ contribution to $N(E_F)$ relative to the Fe-based
superconductors places this material further from magnetism than
those compounds, at least at the LSDA level.
In particular, we found neither a stable ferromagnetic solution,
nor a stable checkerboard antiferromagnetic solution.
However, it should be noted that there is a strong peak in the
DOS $\sim$ 0.1 eV above $E_F$ so that non-stoichiometry
that results in electron doping would be expected to result in ferromagnetism.
Also, in the absence of spectroscopic or other experimental
data for KCo$_2$As$_2$,
the possibility that correlation effects beyond the LSDA could lead
to magnetism cannot be excluded.
In any case,
this is in contrast to our results for KFe$_2$As$_2$, which show a
weak magnetic ground state, ferromagnetic in character, with moments
of 1.1 $\mu_B$/Fe but energy only 13 meV/formula unit lower than
the non-spin-polarized case in the LSDA.

\section{Fe-Co alloys}

\begin{figure}[tbp]
\vspace{0.25cm}
\includegraphics[height=0.96\columnwidth,angle=270]{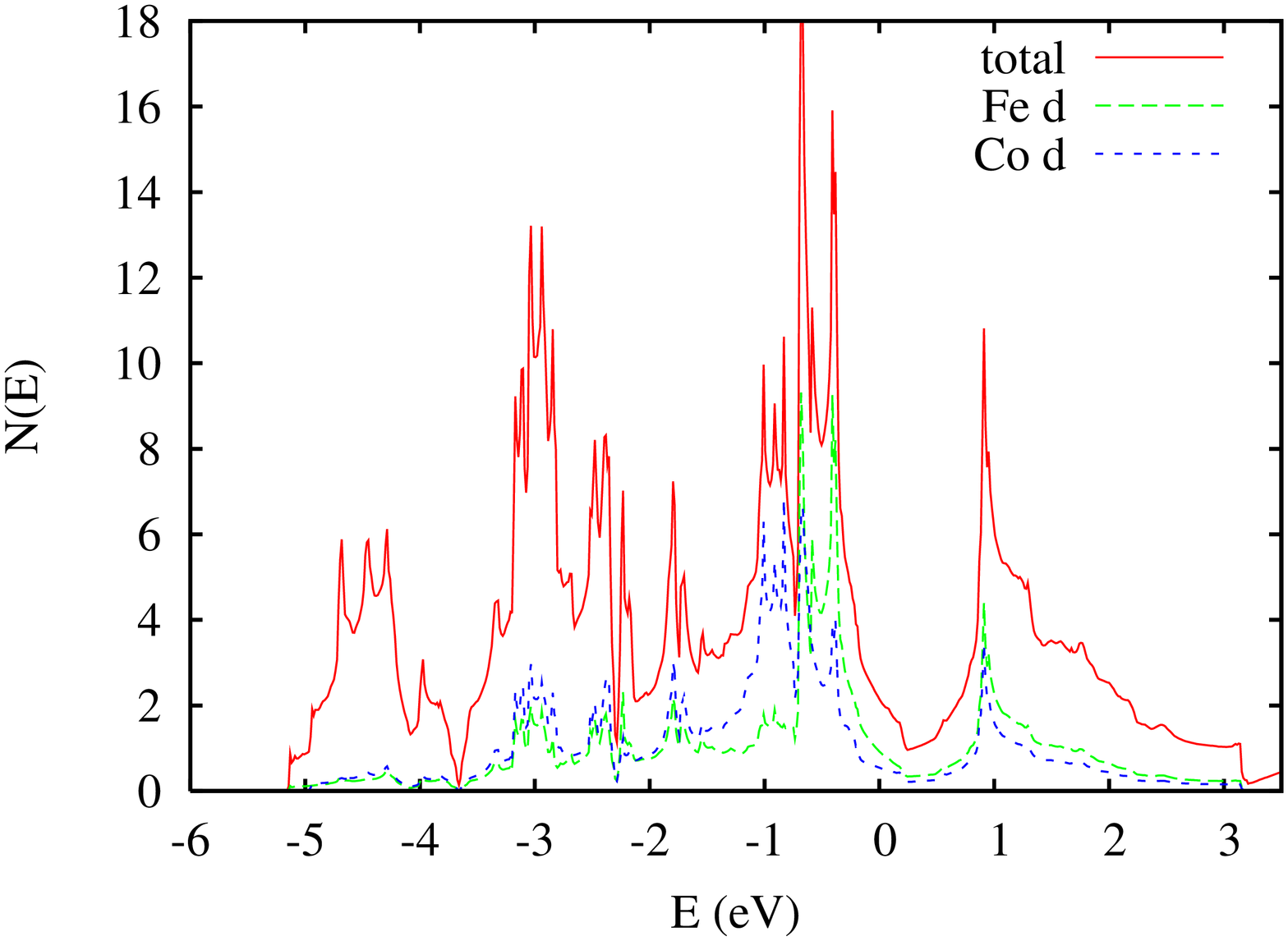}
\includegraphics[height=0.96\columnwidth,angle=270]{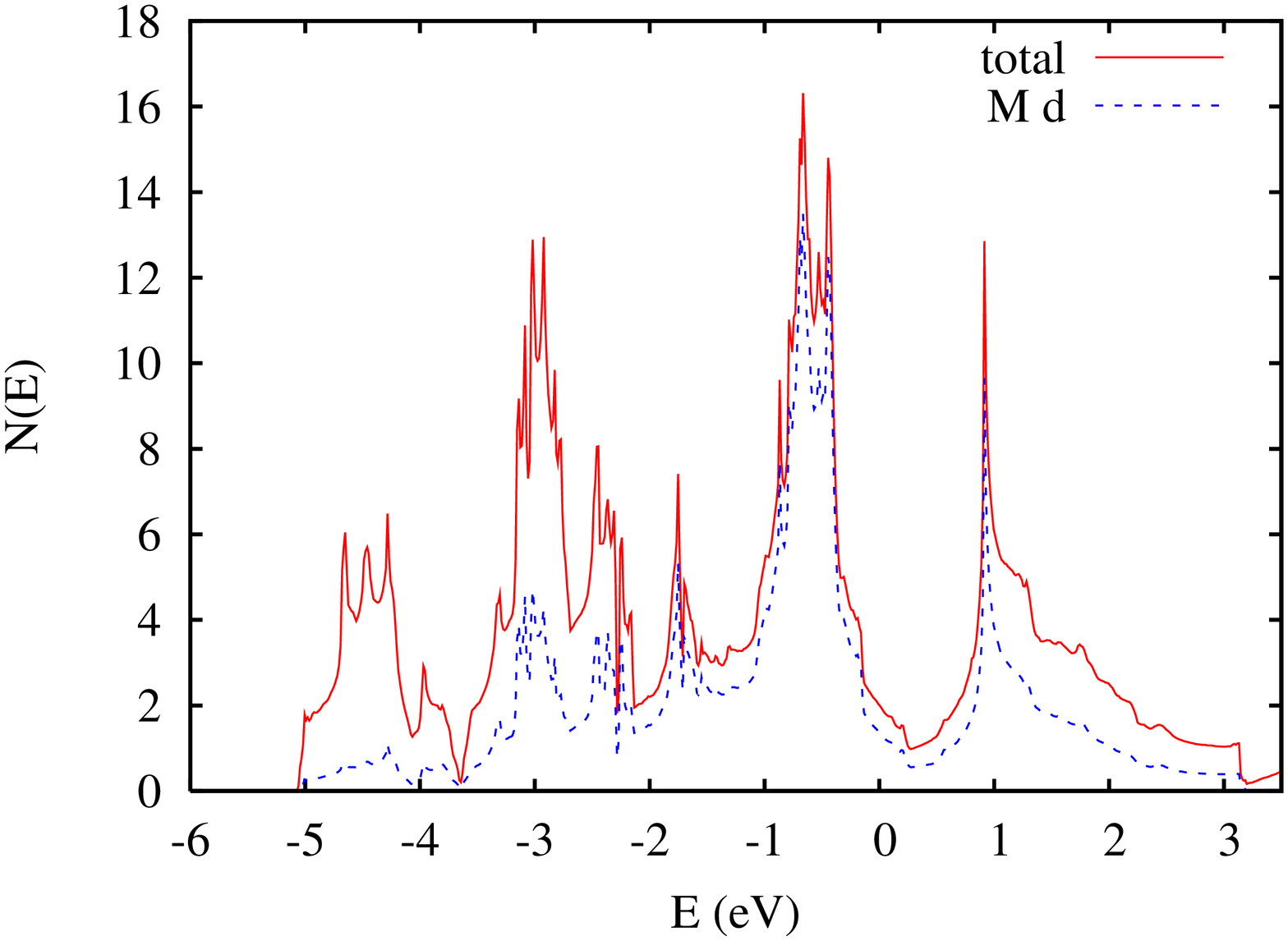}
\caption{(Color online)
Calculated density of states for KFeCoAs$_2$ as obtained
in an ordered cell (top) and in the virtual crystal approximation (bottom).}
\label{dos-ss}
\end{figure}

\begin{figure}[tbp]
\vspace{0.25cm}
\includegraphics[width=0.96\columnwidth,angle=0]{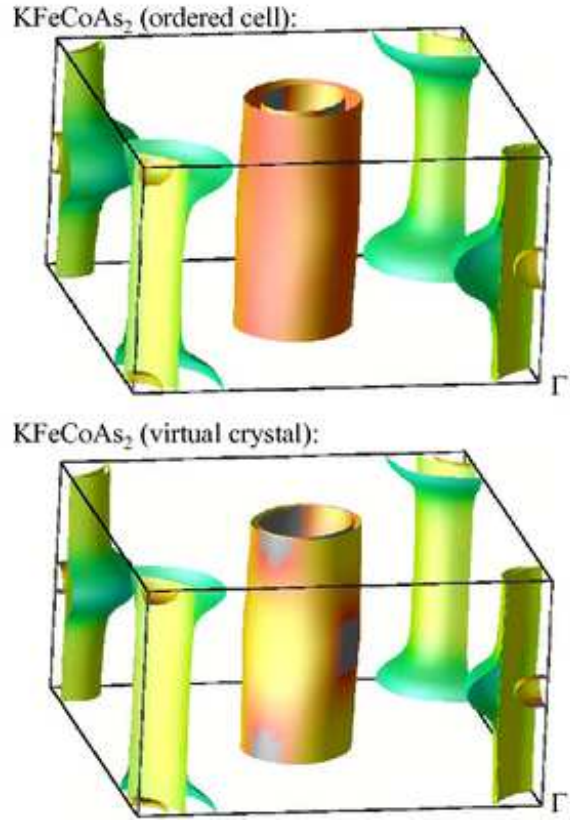}
\caption{(Color online)
Comparison of the Fermi surface of KFeCoAs$_2$ from an ordered
cell and in the virtual crystal approximation.}
\label{fs-comp}
\end{figure}

Considering the similarity of the electronic structure to that of
BaFe$_2$As$_2$ with a Fermi energy shift, we next consider an
alloy of composition KFeCoAs$_2$ (i.e. $y$=1).
This composition has the same valence electron count as BaFe$_2$As$_2$.
Fig. \ref{dos-ss} shows the calculated DOS
for KFeCoAs$_2$ as obtained with a checkerboard ordering of Fe and Co
and as obtained in the virtual crystal approximation.
A comparison of the ordered cell and virtual crystal Fermi surfaces
is given in Fig. \ref{fs-comp}.
In the ordered cell, the Fe $d$ and Co $d$ projections of the
DOS are very similar in shape, and furthermore the results of the
virtual crystal calculation are very similar to those of the ordered
cell. Similarly, the virtual crystal Fermi surface is very much
the same as that of the ordered cell.
The implication is that Fe and Co form an alloy with a coherent
electronic structure in this system, similar to the Ba(Fe,Co)$_2$As$_2$
system, and that we can use the virtual crystal approximation to study
the electronic structure for other values of the composition $y$.
A key point is that the Fermi surface and electronic structure are
rather similar to those of BaFe$_2$As$_2$, and in particular that it
shows disconnected, nested hole and electron Fermi surfaces. The implication
is that KFe$_y$Co$_{2-y}$As$_2$
with $y \sim$ 1 is similar to BaFe$_2$As$_2$.
Therefore it will be interesting to determine if this material is
superconducting either under ambient conditions or with pressure or doping.

\begin{figure}[tbp]
\vspace{0.25cm}
\includegraphics[width=0.96\columnwidth,angle=0]{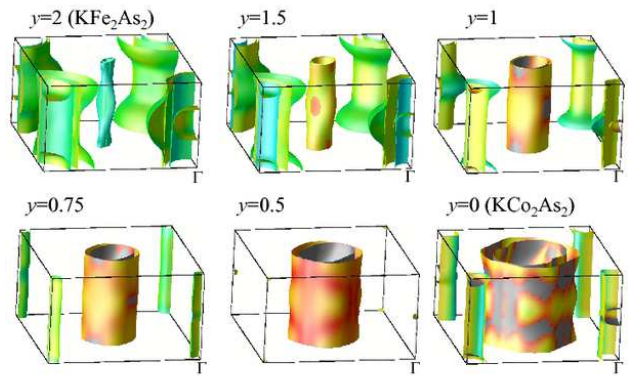}
\caption{(Color online)
Calculated Fermi surface of KFe$_{y}$Co$_{2-y}$As$_2$ for various values
of $y$. Note that the Fermi surface sections around the zone center for
KCo$_2$As$_2$ are different in nature from those at the other
compositions; in particular they are electron sections, and not the
$d_{xz}$,$d_{yz}$ hole sections.}
\label{fs-prog}
\end{figure}

\begin{figure}[tbp]
\vspace{0.25cm}
\includegraphics[height=0.96\columnwidth,angle=270]{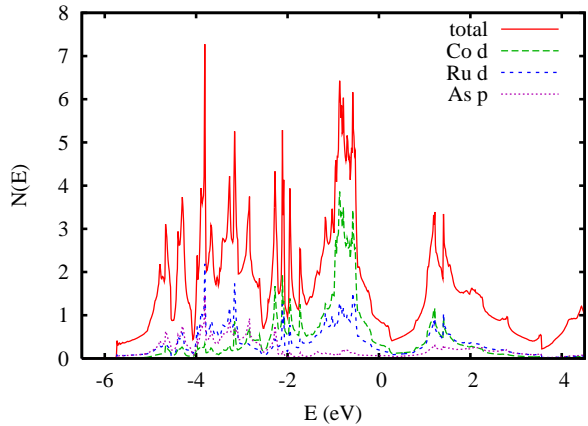}
\caption{(Color online)
Calculated DOS for ordered KRuCoAs$_2$.
}
\label{dos-ru}
\end{figure}

\begin{figure}[tbp]
\vspace{0.25cm}
\includegraphics[width=0.85\columnwidth,angle=0]{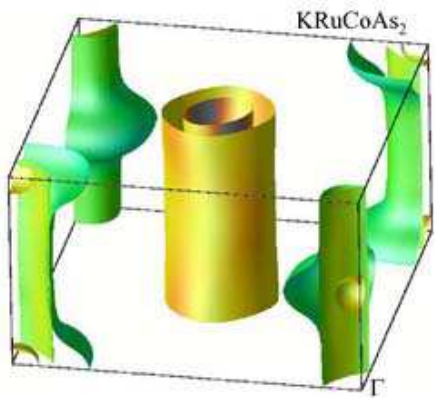}
\caption{(Color online)
Fermi surface for ordered KRuCoAs$_2$.
}
\label{fermi-ru}
\end{figure}

Fig. \ref{fs-prog} shows the evolution of the Fermi surface
of KFe$_{y}$Co$_{2-y}$As$_2$ as a function of $y$, as obtained
in the virtual crystal approximation. As may be seen, the behavior of the
Fermi surface is different for electron and hole doping away from
$y$=1. In particular, the characteristic of having disconnected
hole and electron Fermi surfaces persists all the way to $y$=2 on the
hole doped size, but only up to $\sim$ $y$=3/4 on the electron doped side.
This reflects the fact that the bands making up the electron Fermi surfaces
at the zone corner are lighter than those forming the hole sections.
The near vanishing of the electron Fermi surface for KFe$_2$As$_2$ is
consistent with photoemission results. \cite{sato}

One difference from the pure Fe compounds
(e.g. BaFe$_2$As$_2$) is that we find less tendency towards
magnetism, and in fact at $y$=1 we do not find an SDW ground state.
This may be a result of the fact that the in plane lattice parameter
is smaller in the Co compounds,
e.g. $a$=3.794 \AA, in KCo$_2$As$_2$,
vs. 3.9625 \AA, in BaFe$_2$As$_2$.
This leads to a reduction in the value of $N(E_F)$,
e.g. $N(E_F)$=2.2 eV$^{-1}$ per formula unit (two transition
metals) for ordered KFeCoAs$_2$
vs. 3.0 eV$^{-1}$ for non-spin polarized BaFe$_2$As$_2$ calculated the
same way. \cite{singh-bfa}

In any case, the above results indicate that superconductivity due to
disconnected nested hole and electron Fermi surfaces is possible
for Fe concentrations greater than $y/2$ $\sim$ 0.4, i.e. including a range
of concentrations below the percolation limit for Fe on a square lattice.
One caveat is that there is considerable evidence linking Fe-based
superconductivity with magnetism, and these Co compounds are less
magnetic than the Fe-based materials.
Discovery of superconductivity in this system would therefore imply that
local physics on the Fe-site is not an essential ingredient in Fe-based
superconductivity.

\section{Co-Ru alloys}

Finally, we did calculations for an ordered cell KRuFeAs$_2$
in order to determine whether the characteristic electronic
structure of the Fe-based superconductors is maintained
for alloys involving Co and Ru.
These calculations were motivated by the
recent demonstration that superconductivity can
be induced in Fe-based materials by alloying with Ru.
\cite{paulraj}
In fact, we find that this structure is maintained.
The DOS and Fermi surface are shown in Figs.
\ref{dos-ru} and \ref{fermi-ru}, respectively.
However, one may note that the Ru $d$ states are
more strongly hybridized with As $p$ than the Co states -- a fact that
would work against magnetism, in addition to the already reduced tendency
towards magnetism resulting from the
more extended orbitals in 4$d$ compounds as compared to 3$d$ compounds
(i.e. weaker Hund's coupling).

\section{summary and conclusion}

In summary, we show that the electronic structure
of Co-rich K(Fe,Co)$_2$As$_2$ is similar to those of the
Fe-based superconductors, in particular Fe-rich Ba(Fe,Co)$_2$As$_2$,
when the electron count is similar, and that K(Ru,Co)$_2$As$_2$
also shows similar features, but is less magnetic. These results suggest
that these systems should be explored for possible superconductivity
or magnetism associated with the Fermi surface nesting. Should an
SDW be found, pressure experiments, again searching for superconductivity
would be desirable. While the air sensitivity of the K-based ThCr$_2$Si$_2$
arsenides will no doubt complicate such measurements, an observation of
superconductivity in these systems would be of value as it would
impose significant constraints on the model for superconductivity in the
Fe-based materials, in particular by showing that Fe and
local correlation effects associated with Fe are not essential.

\acknowledgments

This work was supported by the Department of Energy,
Division of Materials Sciences and Engineering.

\bibliography{KCo2As2}
\end{document}